\documentclass[aps, pre, superscriptaddress, twocolumn, tightenlines, showpacs, floatfix, pdftex]{revtex4-1}

\usepackage{amsmath, amsfonts, amssymb}
\usepackage{bm,bbm}
\usepackage{mathtools}
\usepackage{amsmath}

\usepackage{afterpackage}
\usepackage{hyperref}
\hypersetup{
    colorlinks,
    linkcolor=blue,
    citecolor=blue,
    filecolor=blue,
    urlcolor=blue
}
\usepackage{multirow}
\usepackage{tikz} 
\usepackage[normalem]{ulem}
\usepackage{wasysym}

\usepackage{float}
\usepackage{graphicx}
\usepackage[caption=false]{subfig} 
\usepackage{array, booktabs}
	\newcolumntype{P}[1]{>{\centering\arraybackslash}p{#1}} 

\newcommand{\corr}[1]{\langle #1\rangle}

\newcommand{\abs}[1]{\vert #1\vert}

\newcolumntype{C}{>{$}c<{$}}
\AtBeginDocument{
\heavyrulewidth=.08em
\lightrulewidth=.05em
\cmidrulewidth=.03em
\belowrulesep=.65ex
\belowbottomsep=0pt
\aboverulesep=.4ex
\abovetopsep=0pt
\cmidrulesep=\doublerulesep
\cmidrulekern=.5em
\defaultaddspace=.5em
}

\begin{document}

\title{Inferring Hidden Symmetries of Exotic Magnets from Detecting \\ Explicit Order Parameters}
\date{\today}

\author{Nihal Rao}
\affiliation{Arnold Sommerfeld Center for Theoretical Physics, University of Munich, Theresienstr. 37, 80333 M\"unchen, Germany}
\affiliation{Munich Center for Quantum Science and Technology (MCQST), Schellingstr. 4, 80799 M\"unchen, Germany}

\author{Ke Liu}
\email{ke.liu@lmu.de}
\affiliation{Arnold Sommerfeld Center for Theoretical Physics, University of Munich, Theresienstr. 37, 80333 M\"unchen, Germany}
\affiliation{Munich Center for Quantum Science and Technology (MCQST), Schellingstr. 4, 80799 M\"unchen, Germany}

\author{Lode Pollet}
\affiliation{Arnold Sommerfeld Center for Theoretical Physics, University of Munich, Theresienstr. 37, 80333 M\"unchen, Germany}
\affiliation{Munich Center for Quantum Science and Technology (MCQST), Schellingstr. 4, 80799 M\"unchen, Germany}
\affiliation{Wilczek Quantum Center, School of Physics and Astronomy, Shanghai Jiao Tong University, Shanghai 200240, China}

\begin{abstract}
An unconventional magnet may be mapped onto a simple ferromagnet by the existence of a high-symmetry point.
Knowledge of conventional ferromagnetic systems may then be carried over to provide insight into more complex orders.
Here we demonstrate how an unsupervised and interpretable machine-learning approach can be used to search for potential high-symmetry points in unconventional magnets without any prior knowledge of the system.
The method is applied to the classical Heisenberg-Kitaev model on a honeycomb lattice, where our machine learns the transformations that manifest its hidden $O(3)$ symmetry, without using data of these high-symmetry points.
Moreover, we clarify that, in contrast to the stripy and zigzag orders, a set of $D_2$ and $D_{2h}$ ordering matrices provides a more complete description of the magnetization in the Heisenberg-Kitaev model.
In addition, our machine also learns the local constraints at the  phase boundaries, which manifest a subdimensional symmetry.
This paper highlights the importance of explicit order parameters to many-body spin systems and the property of interpretability for the physical application of machine-learning techniques.
\end{abstract}

\maketitle

\section{Introduction} \label{sec:intro}
Applications of machine learning in different fields of physics have become ubiquitous and witnessed a dramatic rise in the past few years ~\cite{Carleo19, Carrasquilla20}, ranging from statistical physics~\cite{Zdeborova16, Sohl15}, condensed matter physics~\cite{Carleo17, Wang16, Carrasquilla17}, chemistry and material science~\cite{Nussinov16, Butler18, Morgan20}, to high energy physics~\cite{Guest18, Radovic18, Ntampaka19} and quantum computation~\cite{Schuld18, Biamonte17, Haah17}.
Although studies in the earlier stages have primarily focused on benchmarking algorithms,
many recent developments are moving towards practical tools for solving more complicated and challenging problems.
Instances of these advances include, for example, discovering new classes of wave functions in strongly correlated systems~\cite{Luo19}, improving the accuracy on atoms and small molecules~\cite{Pfau20, Hermann20}, designing efficient algorithms~\cite{Liao19, Carleo19b, Nagai17}, and analyzing experiments~\cite{Zhang19, Torlai19, Bohrdt19, Khatami20}.

Here we explore the potential of using machine-learning techniques to search for hidden symmetries in many-body spin systems. 
Symmetry is at the heart of our understanding of physics. Apparent symmetries such as time, spatial, and rotational invariance lead to the conservation of energy, momentum, and angular momentum, respectively.
However, quite often, the effective symmetry of a system is not apparent, which we henceforth refer to as hidden symmetry.
For instance, in some extended Kitaev systems, which are subject to active research due to their proximity to Kitaev spin liquids (KSLs)~\cite{Kitaev06} and other exotic phases~\cite{Jackeli09, Chaloupka10, Takagi19, Janssen19, Winter17b}, there exist high-symmetry points.
At these points, a complex ordering pattern may be transformed to a simple one~\cite{Chaloupka15, Chaloupka13, Rusnacko19}.
Knowledge of conventional orders can then be carried over, and pseudo-Goldstone modes may be realized even when the Hamiltonian seemingly manifests a low discrete symmetry. 
Others remarkable examples are the Bethe-ansatz solvable $SU(3)$ point in the spin-$1$ bilinear-biquadratic chain~\cite{Uimin70, Lai74, Sutherland75, Batista02} and the emergent $O(4)$ symmetry in the spin-$1/2$ $J$-$Q$ model~\cite{ZhaoSandvik2019}.

Although hidden symmetries are of broad relevance and rich in physics, identifying them is a non-trivial task and is very much problem-dependent, often requiring remarkable insights and experience from researchers.
Therefore, it would be interesting and useful if machine-learning techniques can facilitate their identification.

In this paper, we use a machine-learning method, the tensorial-kernel support vector machine (TK-SVM)~\cite{Greitemann19, Liu19, Greitemann19b}, to find potential hidden symmetries in a spin model.
This method is \emph{interpretable} and \emph{unsupervised}.
The term ``interpretable'' means the machine classifiers can be systematically decoded to physical order parameters~\cite{Greitemann19, Liu19}.
This is crucial in physical applications, as an ultimate goal of learning phase diagrams is to understand the nature of each phase and find suitable characterizations. 
The term ``unsupervised'' means pre-labeled data and prior knowledge of a phase diagram of interest are not required in training, since the supervision of standard support vector machines (SVMs) will be taken over by graph partitioning~\cite{Liu19, Greitemann19b}.

We show that our method provides an efficient and versatile approach to detect high-symmetry points hidden in unconventional magnets.
We demonstrate the method by applying it to the classical Heisenberg-Kitaev (HK) model on a honeycomb lattice, where our machine correctly identifies its hidden $O(3)$ symmetries and the associated transformations.
Moreover, we clarify that the pictorial description of the zigzag and stripy orders only partially reflects the ordering in the HK model. The complete orders are characterized by a set of $D_2$ and $D_{2h}$ ordering matrices.

The paper is organized as follows. In Section~\ref{sec:model_method} we define the HK Hamiltonian and review the TK-SVM method.
Section~\ref{sec:pd} discusses the machine-learned phase diagram.
Section~\ref{sec:op} is devoted to explicit order parameters and the corresponding magnetization curves.
The connection between hidden symmetries and ordering matrices is given in Section~\ref{sec:hidden}.
Section~\ref{sec:constrains} provides a discussion of local constraints and a subdimensional symmetry at phase boundaries. 
We conclude in Section~\ref{sec:conclusion} with an outlook.

\begin{figure}[t]
  \centering
  \includegraphics[width=0.35\textwidth]{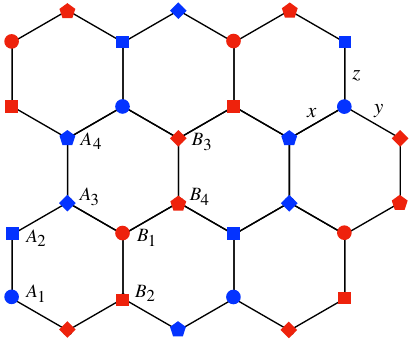}
  \caption{Depiction of the a honeycomb lattice and the $D_{2h}$ and $D_2$ magnetic cell, which contain eight spins and two sectors marked by $A$ (blue) and $B$ (red).
  This choice of magnetic cell fits zigzag and stripy patterns along different directions and also applies to states at the hidden $O(3)$ points which cannot be captured by a four-site zigzag or stripy cell.
   $x,y$, and$z$ label the three distinct bonds in the Kitaev interaction.}
  \label{fig:lattice}
\end{figure}

\section{Model and method} \label{sec:model_method}

We consider the HK model on a honeycomb lattice to demonstrate the concept.
It should be noted however that the following discussion is intended to provide a general guidance for using TK-SVM to search for unconventional orders and hidden symmetries, and is transferable to other spin systems.

\subsection{Heisenberg-Kitaev Hamiltonian} \label{sec:HK_model}
The honeycomb HK model is defined as
\begin{align}\label{eq:model}
	H = \sum_{\corr{ij}_\gamma}J \vec{S}_i \cdot \vec{S}_j  + K S_i^\gamma S_j^\gamma,
\end{align}
where $J$ and $K$ denote the Heisenberg and Kitaev interaction, respectively, and can be parametrized by an angle variable $\varphi \in [0, 2\pi)$ with $K = \sin\varphi$, $J = \cos\varphi$;
$\gamma \in \{x, y, z\}$ labels the three types of nearest-neighbor bonds $\corr{ij}_\gamma$, as depicted in Figure~\ref{fig:lattice}.

The spin-$\frac{1}{2}$ version of the HK model accommodates four magnetic orders and two extended regions of quantum (KSLs) [Chaloupka10, Chaloupka13].
In the large-$S$ limit, the four magnetic orders persistent, while the counterpart classical KSLs only exist at two single points $K = \pm 1$ and $J = 0$ at zero temperature.
Nevertheless, the transformations identifying the hidden symmetry points, which are inside two magnetic phases, are the same.

\subsection{TK-SVM} \label{sec:tksvm}

The TK-SVM is an interpretable and unsupervised approach to detect general symmetry-breaking spin orders~\cite{Greitemann19, Liu19} and emergent local constraints~\cite{Greitemann19b, Liu21}.
It is formulated in terms of the decision function
\begin{align}\label{eq:d(x)}
    d(\mathbf{x}) = \sum_{\mu\nu} C_{\mu\nu} \phi_\mu(\mathbf{x}) \phi_\nu(\mathbf{x}) - \rho.
\end{align}
Here, $\mathbf{x} = \{S_i^a | a = x,y,z; i = 1,2, \dots, N\}$ denotes configurations of $N$ spins and serves as training data.
$\phi_\mu(\mathbf{x})$ maps $\mathbf{x}$ to a tensorial feature space, the $\phi$ space, which can represent general spin orders~\cite{Nissinen16, Michel01}, regardless of exotic magnets, multipolar tensorial orders~\cite{Greitemann19, Liu19} and emergent local constraints~\cite{Greitemann19b, Liu21}.
$C_{\mu\nu}$ can be viewed as an encoder of order parameters, from which explicit expressions of the detected orders are identified.  
$\rho$ is a bias parameter probing whether two sample sets originate from the same phase.
See Appendix~\ref{app:svm} for details.

Although the decision function Eq.~\eqref{eq:d(x)} carries out a binary classification between two sets of data, TK-SVM can also classify multiple data sets.
Such a multiclassification is essentially realized by individual binary problems but makes it possible to compute a phase diagram via unsupervised graph partitioning.

Consider a spin Hamiltonian characterized by a number of physical parameters, such as temperature and different kinds of interactions.
We can cover its parameters space, $\mathcal{V}$, by a grid of the same dimensionality. 
The choice of the grid is arbitrary, either uniform or distorted to have denser nodes in the most interesting subregions of $\mathcal{V}$.
We collect spin configurations $\mathbf{x}$ at vertices of the grid and perform the SVM  multiclassification on the sampled data.
For a grid of $M$ vertices, this will produce $M(M-1)/2$ decision functions as Eq.~\eqref{eq:d(x)}, composed of binary classifications between each pair of vertices.
We then introduce a weighted edge between two vertices, and the weight, $w(\rho) \in [0,1)$, is based on the bias parameter in the corresponding $d(x)$.
In this way, we create a graph with $M$ vertices and $M(M-1)/2$ edges; its partitioning will give the phase diagram.

In formal terms, the graph can be described by a $M\times M$ Laplacian matrix $\hat{L}$.
The off-diagonal entries of $\hat{L}$ accommodate edge weights connecting vertices, and the diagonal entries are degrees of those vertices.
The partitioning can be solved by Fiedler's theory of spectral clustering~\cite{Fiedler73, Fiedler75},
\begin{align}\label{eq:L_f}
    \hat{L} \mathbf{f}_i = \lambda_i \mathbf{f}_i.
\end{align}
As $\hat{L}$ is positive semi-defined, the smallest possible eigenvalue is $\lambda_1 = 0$, corresponding to a trivial eigenvector $(1 \, 1\, \dots \, 1)^{\rm T}$.
The second smallest eigenvalue $\lambda_2$ measures the algebraic connectivity of the graph~\cite{Fiedler73, Fiedler75}. The corresponding eigenvector $\mathbf{f}_2$ is referred to as the \textit{Fiedler vector}, which reflects how the vertices are clustered and plays the role of a phase diagram in the context of TK-SVM~\cite{Liu19, Greitemann19b}.
We refer to Appendix~\ref{app:graph} for details.

\begin{figure*}[t!]
    \centering
    \includegraphics[width=0.8\textwidth]{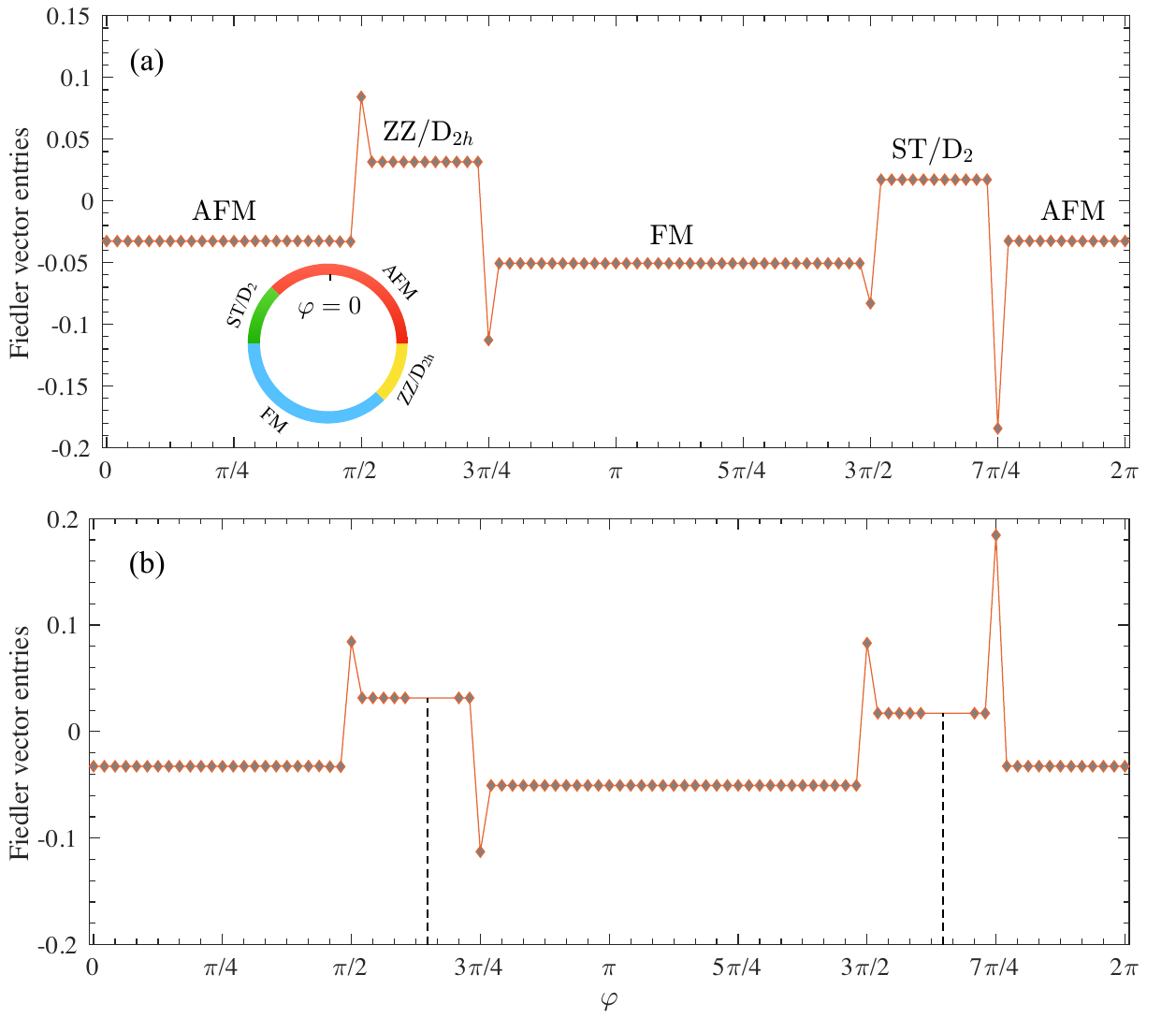}
     \caption{The Fiedler vector acts as the phase diagram of the Heisenberg-Kitaev model.
     (a) Gradients in Fiedler vector entries reflect the clustering of the graph. The plateaus indicate stable phases, and the jumps signal phase transitions. The phases are interpreted in Section~\ref{sec:op} and are labeled following the common convention: AFM, antiferromagnet; ZZ, zigzag; FM, ferromagnet; ST, stripy.
     In addition, the ST and ZZ region are also marked according to the $D_2$ and $D_{2h}$ magnetization measured in Figure~\ref{fig:measure}. The inner panel shows a circular representation of the phase diagram.
      (b) Another partitioning with removing data near the high-symmetry points $\varphi = \arctan(-2) \sim 0.65 \pi$ and $1.65 \pi$ (indicated by the dashed lines; the graph is not shown), to demonstrate that data of these special points are not needed for revealing the hidden $O(3)$ symmetry.
      The partitioning is reflected by contrasts between Fiedler vector entries, rather than the absolute values. Panels (a) and (b) lead to the same topology of the phase diagram.}
    \label{fig:pd}
\end{figure*}

\section{Machine-learned phase diagram} \label{sec:pd}

A typical application of TK-SVM consists of two steps: (i) detecting the topology of the phase diagram and (ii) extracting and verifying order parameters.
We focus here on the classical phase diagram of the HK model Eq.~\eqref{eq:model}, and save the discussion of order parameters for the next section.

For this purpose, we introduce a fictitious grid that spans uniformly in the space of $\varphi$, with a spacing of $\delta\varphi = \frac{\pi}{48}$.
At each $\varphi$, we collect $500$ spin configurations at a low temperature $T = 10^{-3}\sqrt{J^2+K^2}$.
The samples are prepared by classical parallel tempering Monte Carlo simulations on a lattice of $10,368$ spins ($72 \times 72$ honeycomb unit cells).
Next, we perform TK-SVM with different ranks over these data.
However, it turns out that a rank-$1$ TK-SVM (see Appendix~\ref{app:svm}), which detects magnetic orders, is sufficient to learn the phase diagram.
The result is a graph of $96$ vertices and $4,560$ edges.
The Fiedler vector obtained from partitioning the graph is depicted in Figure~\ref{fig:pd} (a) (see also Appendix~\ref{app:graph}).
Each of its entries represents a vertex of the grid, hence a $\varphi$-point.
The Fiedler vector entries for the vertices ($\varphi$s) classified in the same subgraph component are identical or very close in value, while those falling into different subgraphs display considerable contrast. 

Evidently, the Fiedler vector shows four subgraph components, indicating four stable phases.
This in fact reproduces the classical HK phase diagram~\cite{Price13, Janssen16}.
The four plateaus respectively correspond to the antiferromagnetic (AFM), zigzag (ZZ), ferromagnetic (FM) and stripy (ST) phase, following the labeling in Figure~\ref{fig:pd} (a).
However, as we shall discuss in Section~\ref{sec:op}, orders in the regions
$\varphi \in (\frac{\pi}{2}, \frac{3\pi}{4})$ and $(\frac{3\pi}{2}, \frac{7\pi}{4})$ may be more universally measured by $D_2$ and $D_{2h}$ magnetization.

\begin{table}[t]
\centering
\renewcommand{\arraystretch}{2}
\newcolumntype{C}[1]{>{\centering\arraybackslash$}m{#1}<{$}}
\newlength{\mycolwd}
\newcommand\SmallMatrix[1]{\settowidth{\mycolwd}{$a-1$}\scalebox{0.618}{\renewcommand{\arraystretch}{0.85}\ensuremath{\left(\begin{array}{*{3}{@{}C{\mycolwd}@{}}} \displaystyle #1 \end{array}\right)}}}
\begin{tabular}{l l l}
   \hline
   \hline
    {\bf Phases} \hspace{0.2cm} & {\bf Ordering Matrices} &  \\
    \midrule
   $\mathrm{D_2}$ 
      	& $\hat{T}^{\scalebox{0.618}{A,B}}_1 = \SmallMatrix{
            1 & 0 & 0 \\
            0 & 1 & 0 \\
            0 & 0 & 1
        }$, 
        & $\hat{T}^{\scalebox{0.618}{A,B}}_2 =  \SmallMatrix{
            -1 & 0 & 0 \\
            0 & -1 & 0 \\
            0 & 0 & 1
        }$,
        \\[5pt]
        & $\hat{T}^{\scalebox{0.618}{A,B}}_3 = \SmallMatrix{
            -1 & 0 & 0 \\
            0 & 1 & 0 \\
            0 & 0 & -1
        }$, 
        & $\hat{T}^{\scalebox{0.618}{A,B}}_4 =  \SmallMatrix{
            1 & 0 & 0 \\
            0 & -1 & 0 \\
            0 & 0 & -1
        }$
  \\[5pt]
    \hline
    $\mathrm{D_{2h}}$  
      	& $\hat{T}^{\scalebox{0.618}{A,B}}_1 = \pm \SmallMatrix{
            1 & 0 & 0 \\
            0 & 1 & 0 \\
            0 & 0 & 1
        }$,
        & $\hat{T}^{\scalebox{0.618}{A,B}}_2 =  \pm \SmallMatrix{
            1 & 0 & 0 \\
            0 & 1 & 0 \\
            0 & 0 & -1
        }$,
        \\[5pt]
        & $\hat{T}^{\scalebox{0.618}{A,B}}_3 = \pm \SmallMatrix{
            -1 & 0 & 0 \\
            0 & 1 & 0 \\
            0 & 0 & -1
        }$,
        & $\hat{T}^{\scalebox{0.618}{A,B}}_4 =  \pm \SmallMatrix{
            -1 & 0 & 0 \\
            0 & 1 & 0 \\
            0 & 0 & 1
        }$
	\\[5pt]
  \hline
  \hline
\end{tabular}
\caption{$D_2$ and $D_{2h}$ ordering matrices. Their magnetic cells are shown in Figure~\ref{fig:lattice}, which consist of two sectors, labeled by $A, \, B$, and in total eight sublattices.
The $D_2$ and $D_{2h}$ orders involve four and eight distinct spin orientations, respectively, and are described by the respective three-dimensional dihedral groups.
Their ordering matrices also define the sublattice transformations that identify the hidden $O(3)$ points in the Heisenberg-Kitaev model.}\label{tab:ops}
\end{table}

Sudden jumps in the Fiedler vector entries manifest phase transitions, which are seen to occur at 
$J = 0, \, K = \pm 1$ ($\varphi = \frac{\pi}{2}, \, \frac{3\pi}{2}$) and
$J = -K$ ($\varphi = \frac{3\pi}{4}, \, \frac{7\pi}{4}$). 
The boundaries at $K = \pm 1$ correspond to the Kitaev limits.
Different from the cases of quantum spin-$\frac{1}{2}$ and spin-$1$, where KSLs are proposed to extend to finite regions of $J$~\cite{Chaloupka10, Chaloupka13, Osorio14, Gohlke17, Wang19, Dong20}, in the large-$S$ limit, KSLs are unstable against the Heisenberg interaction and reduce to critical points.
Nevertheless, this will not affect our discussion of the hidden symmetries in the HK model.

We note that the learning of Figure~\ref{fig:pd} is unsupervised. No prior knowledge of the phase diagram and order parameters was used, and all the four phases are discriminated simultaneously by a \emph{single} partitioning.
Moreover, after we determine the global topology of the phase diagram, the resolution of which is set by the given training dataset, phase boundaries can be further refined by directly examining the learned order parameters.

\begin{figure*}[t]
    \centering
	\includegraphics[width=0.85\textwidth]{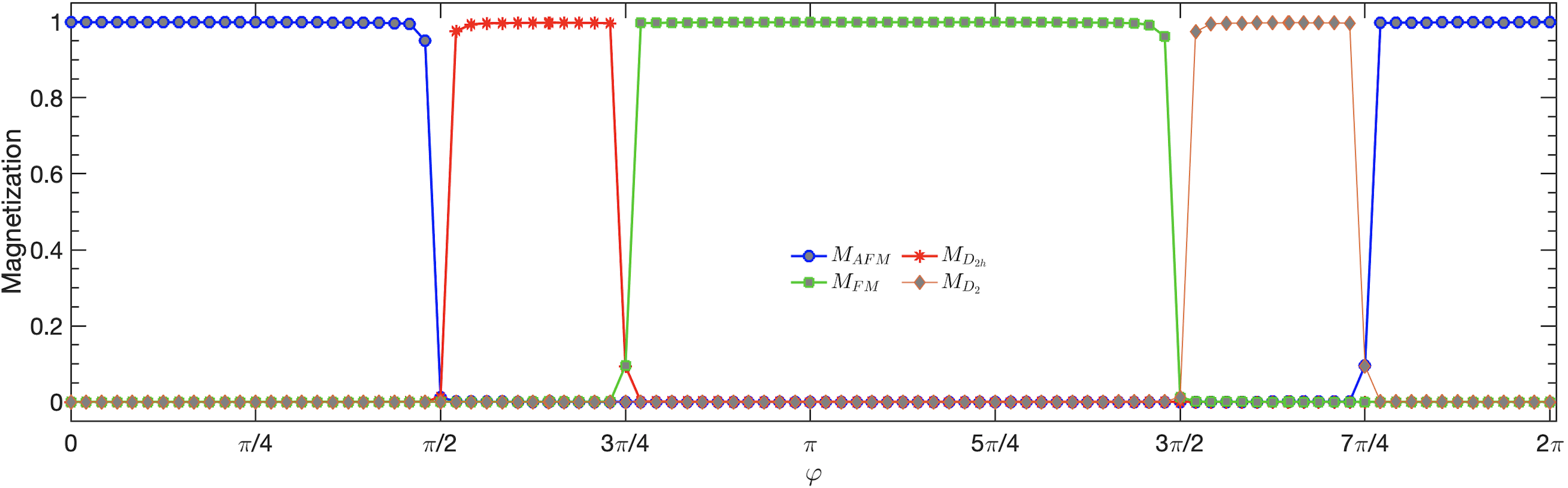}
    \caption{Measurements of order parameters. The FM, AFM, $D_2$ and $D_{2h}$ magnetization are measured as a function of $\varphi$ at low temperature $T = 10^{-3} \sqrt{J^2+K^2}$.
    In each phase, the respective magnetization saturates to unity ($M = \big\langle\abs{\frac{1}{N_{\rm cell}}\sum_{\rm cell}\protect\overrightarrow{M}}\big\rangle =1$), while others vanish, where $\protect\overrightarrow{M}$ is the ordering moment in one magnetic cell, $\sum_{\rm cell}$ sums over magnetic cells, and $\langle \dots \rangle$ denotes the ensemble average.
    The small residual moments at $\varphi = 0.75 \pi$ and $1.75 \pi$ are finite-size effects. At these points, the classical ground states form decoupled FM and AFM Ising chains with a subextensive degeneracy (Section~\ref{sec:constrains}).}  
    \label{fig:measure}
\end{figure*}

\begin{figure}[t]
    \centering
    \includegraphics[width=0.48\textwidth]{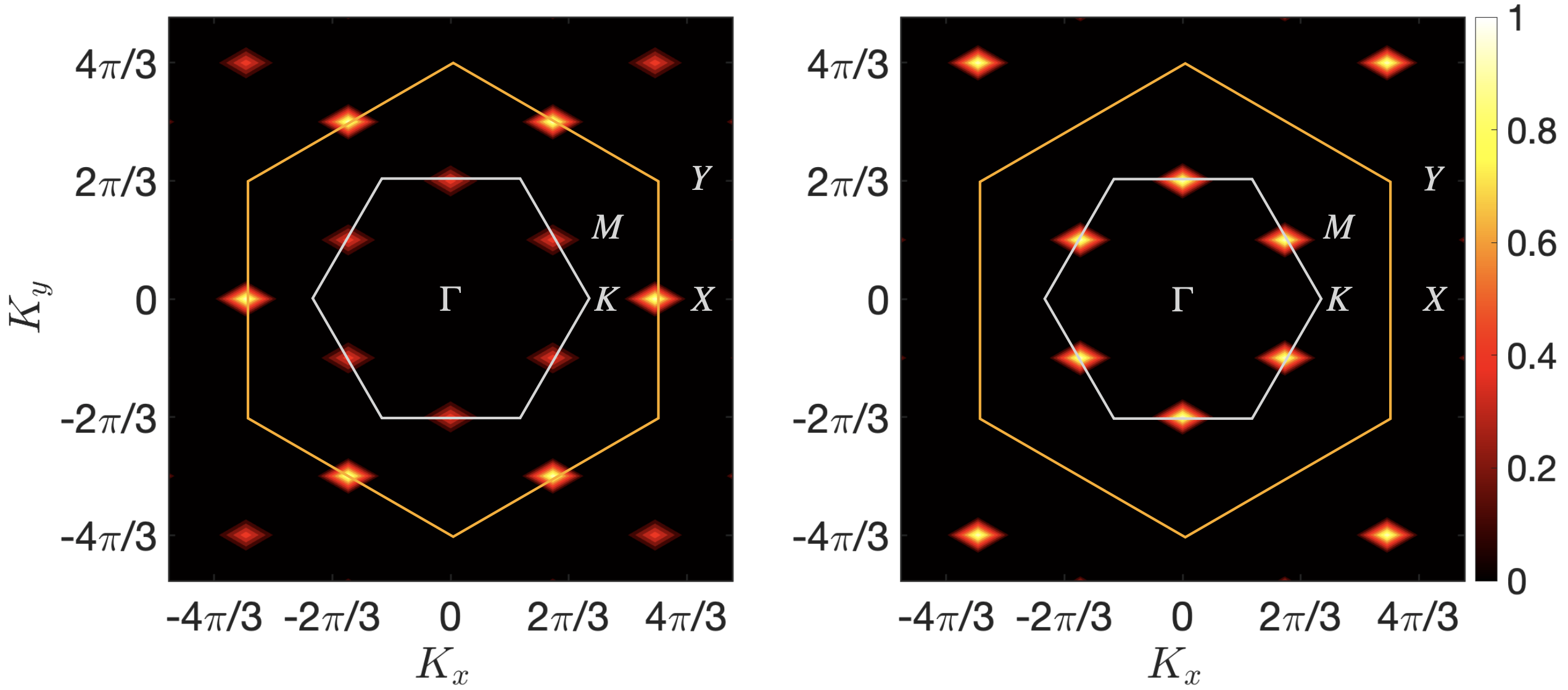}
    \caption{Static spin-structure factor, $S (\vec{K})$, for the ST and $D_{2}$ (left) and ZZ and $D_{2h}$ order (right). 
    The orange and gray hexagon denote the second and first honeycomb Brillouin zone respectively, and high-symmetry points are indicated.
  $S (\vec{K}) =  \big\langle \frac{1}{N} \sum_{ij} \vec{S}_i \cdot \vec{S}_j \, e^{i \vec{K} \cdot (\vec{r}_i - \vec{r}_j)}
  \big\rangle$, where $\vec{r}_i$ is the position of a spin at site $i$, and a nearest-neighbor bond of the honeycomb lattice is set to unit length.}
    \label{fig:SSF}
\end{figure}

\begin{figure}[t]
    \centering
    \begin{tabular}{c}
    	\subfloat[$D_2$ pattern]{\includegraphics[width=0.35\textwidth]{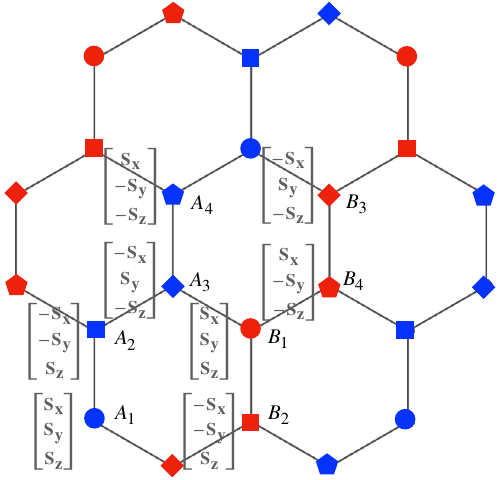}} \\
    	\subfloat[$D_{2h}$ pattern]{\includegraphics[width=0.35\textwidth]{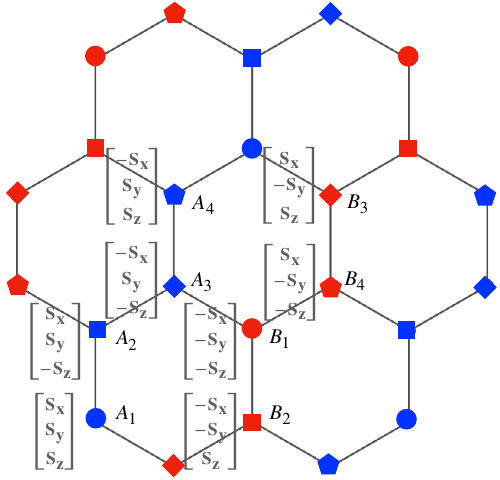}}
    \end{tabular}
    \caption{Configurations of an arbitrary $D_2$ and $D_{2h}$ state. The spin $\vec{S}_{A_1} = (S_x \ S_y \ S_z)^{\rm T}$ is used as the reference spin, while orientations of other spins are determined according to the respective ordering matrices.
    Compared to stripy and zigzag orders, which are staggered arrangements of $\pm \vec{S}$, the sign flip in a $D_{2}$ and $D_{2h}$  pattern can occur at individual components. 
     In special cases $\vec{S}_{A_1} = (0 \ 0 \,\pm\! 1)^{\rm T}$, these patterns are equivalent to the $Z$-type zigzag and stripy patterns in Figure~\ref{fig:ST_ZZ}, with a reduced four-site magnetic cell $\{A_1, A_2, A_3, B_3\}$.
    When choosing $\vec{S}_{A_1} = (\pm 1 \ 0 \ 0)^{\rm T}$ and $(0 \, \pm\! 1 \ 0)^{\rm T}$, $X$- and $Y$-type zigzag and stripy states will be realized, where the magnetic cells are given by  $\{A_1, A_2, A_3, B_1\}$ and  $\{A_2, A_3, B_1, B_2\}$, respectively. In general cases, the $D_{2h}$ ($D_2$) and zigzag (stripy) orders are different, and the magnetic cell cannot be reduced to four sites.}
    \label{fig:D2_D2h}
\end{figure}

\begin{figure}[ht!]
    \centering
    \begin{tabular}{cc}
    	\subfloat[Stripy]{\includegraphics[width=0.2\textwidth]{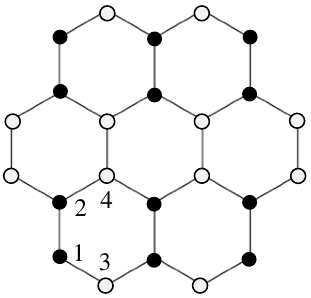}} &
    	\subfloat[Zigzag]{\includegraphics[width=0.2\textwidth]{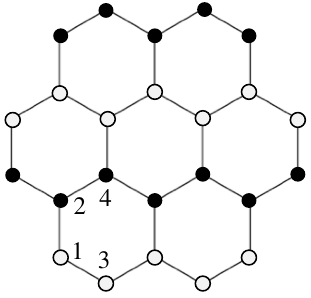}}
    \end{tabular}
    \caption{Representative configurations of a stripy (ST) and zigzag (ZZ) order. White ($\vec{S}$) and black ($-\vec{S}$) cycles denote opposite spins.
    The corresponding magnetization can be defined as 
    $M_{\rm ST} = \big\langle\abs{\frac{1}{N_{\rm cell}}\sum_{\rm cell} (\vec{S}_1 + \vec{S}_2 - \vec{S}_3 - \vec{S}_4)}\big\rangle$,
    and $M_{\rm ZZ} = \big\langle\abs{\frac{1}{N_{\rm cell}}\sum_{\rm cell} (\vec{S}_1 - \vec{S}_2 + \vec{S}_3 - \vec{S}_4)}\big\rangle$, respectively, where the numbers label the four sublattices.
    In general, $\vec{S}$ may point to arbitrary directions. However, in the ground states of the Heisenberg-Kitaev model, the realization of these above configurations will be accompanied by $\vec{S} = (0 \ 0 \,\pm\! 1)^{\rm T}$. We hence refer to them as $Z$ type.
    Such states are present in the intersection of zigzag (stripy) and $D_{2h}$ ($D_2$) manifolds.}
    \label{fig:ST_ZZ}
\end{figure}

\begin{figure}[ht!]
    \centering
    \includegraphics[width=0.5\textwidth]{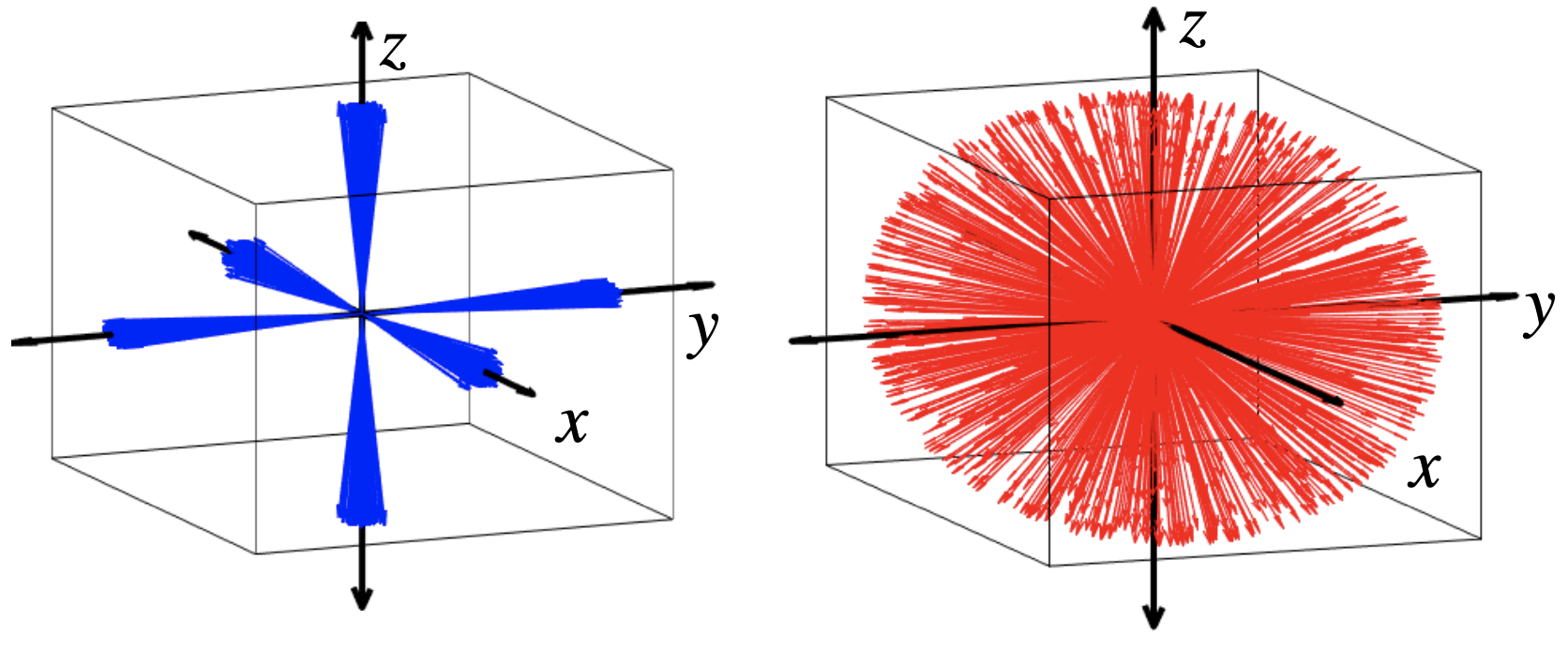}
    \caption{Distribution of spin orientations for states
    in the ZZ or $D_{2h}$ and ST or $D_{2}$ phases away (left) and at (right) the hidden $O(3)$ points, at a low temperature $T=0.001$.}
    \label{fig:spins}
\end{figure}

\section{Explicit order parameters} \label{sec:op}
We move on to interpret the nature of the phases shown in the  phase diagram of  Figure~\ref{fig:pd}. 
By virtue of the strong interpretability, analytical order parameters can be extracted from the corresponding $C_{\mu\nu}$ matrix (see Appendix~\ref{app:svm}).
As the FM and AFM orders are trivial, we will focus on the other two phases.

The $D_2$ and $D_{2h}$ magnetization can be expressed as
\begin{align}\label{eq:op}
    \overrightarrow{M} = \frac{1}{8} \sum_{\scalebox{0.618}{A,B}}\sum_{k = 1}^{4} \hat{T}^{\scalebox{0.618}{A,B}}_{k}\vec{S}_{k}.
\end{align}
Here the ordering matrices $\hat{T}^{\scalebox{0.618}{A,B}}_{k}$ describe the relative orientation of spins in a magnetic cell and are tabulated in Table~\ref{tab:ops}. The subscript $k$ has a slightly different numeration in the two sublattice sectors  $A$ and $B$ as illustrated in Figure~\ref{fig:lattice} (also Figure~\ref{fig:D2_D2h}).

The $D_2$ order is formulated by four different matrices with
 $\hat{T}^{\scalebox{0.618}{A}}_{k} = \hat{T}^{\scalebox{0.618}{B}}_{k}$,
 forming the three-dimensional dihedral group $D_2$.
These matrices have been proposed in the study of orbital degeneracy of Mott insulators~\cite{Khaliullin02, Khaliullin05} and are used to identify the hidden symmetries of the HK model~\cite{Chaloupka10, Chaloupka15}, which will be discussed in Section~\ref{sec:hidden}.
The $D_{2h}$ order can be viewed as an AFM version of the $D_2$ order, where
$\hat{T}^{\scalebox{0.618}{A}}_{k} = -\hat{T}^{\scalebox{0.618}{B}}_{k}$
in the respective sublattice.
It is thereby aptly named after the dihedral group $D_{2h} \cong D_2 \times Z_2$.

These order parameters, as well as the FM and Neel orders, are measured at $T = 10^{-3}\sqrt{J^2 + K^2}$ which is the temperature during training the TK-SVM.
As shown in Figure~\ref{fig:measure}, the respective magnetization saturates to unity,  spans the entire phase, and vanishes in the other phases.

The measurements of $D_{2h}$ and $D_{2}$ magnetization validate that they are the correct order parameters for the regions
$\varphi \in (\frac{\pi}{2}, \frac{3\pi}{4})$ and $(\frac{3\pi}{2}, \frac{7\pi}{4})$. 
These regions are traditionally described by zigzag and stripy orders~\cite{Price12, Price13, Janssen16}, which have the same static structure factor as the $D_{2h}$ and $D_{2}$ order as shown in Figure~\ref{fig:SSF}.
We now discuss the relation and differences between these orders.
 
Figure~\ref{fig:D2_D2h} shows configurations of a $D_{2h}$ and a $D_2$ state, which can be generated by fixing one spin, e.g. $\vec{S}_{A_1} = \vec{S}_0$, and determining the orientation of other spins according to the respective ordering matrices in Table~\ref{tab:ops}.
In general, the reference spin $\vec{S}_0$ may point along any direction.
However, there are special instances where the $D_{2h}$ and $D_2$ structures can reduce to the zigzag and stripy orders respectively.
For example, the case $\vec{S}_{0} = (0 \ 0 \,\pm\! 1)^{\rm T}$ reduces to the $Z$-type zigzag and stripy state as shown in Figure~\ref{fig:ST_ZZ}.
Similarly, choosing $\vec{S}_{0} = (\pm 1 \ 0 \ 0)^{\rm T}$ and $(0 \, \pm\! 1 \ 0)^{\rm T}$ will lead to an $X$- and $Y$-type zigzag and stripy state, respectively.
Namely, the manifolds of the zigzag (stripy) and $D_{2h}$ ($D_{2}$) order have overlaps.

In the ZZ/$D_{2h}$ and ST/$D_{2}$ regions, away from the hidden symmetry ($O(3)$) points at $\varphi \approx 0.65 \pi, 1.65 \pi$, the above special states are realized as the ground states of the HK model owing to the discrete symmetry of the Kitaev term; as visualized in Figure~\ref{fig:spins}.
For these states, the distinction between $D_{2h}$ ($D_{2}$) and zigzag (stripy) orders is superfluous.
Nevertheless, once spins are unlocked from the axes, which happen at the $O(3)$ points, the states can no longer be described by staggered arrangements of $\pm \vec{S}$ as in a zigzag or stripy structure.

\begin{figure}[ht!]
  \centering
  \includegraphics[width=0.45\textwidth]{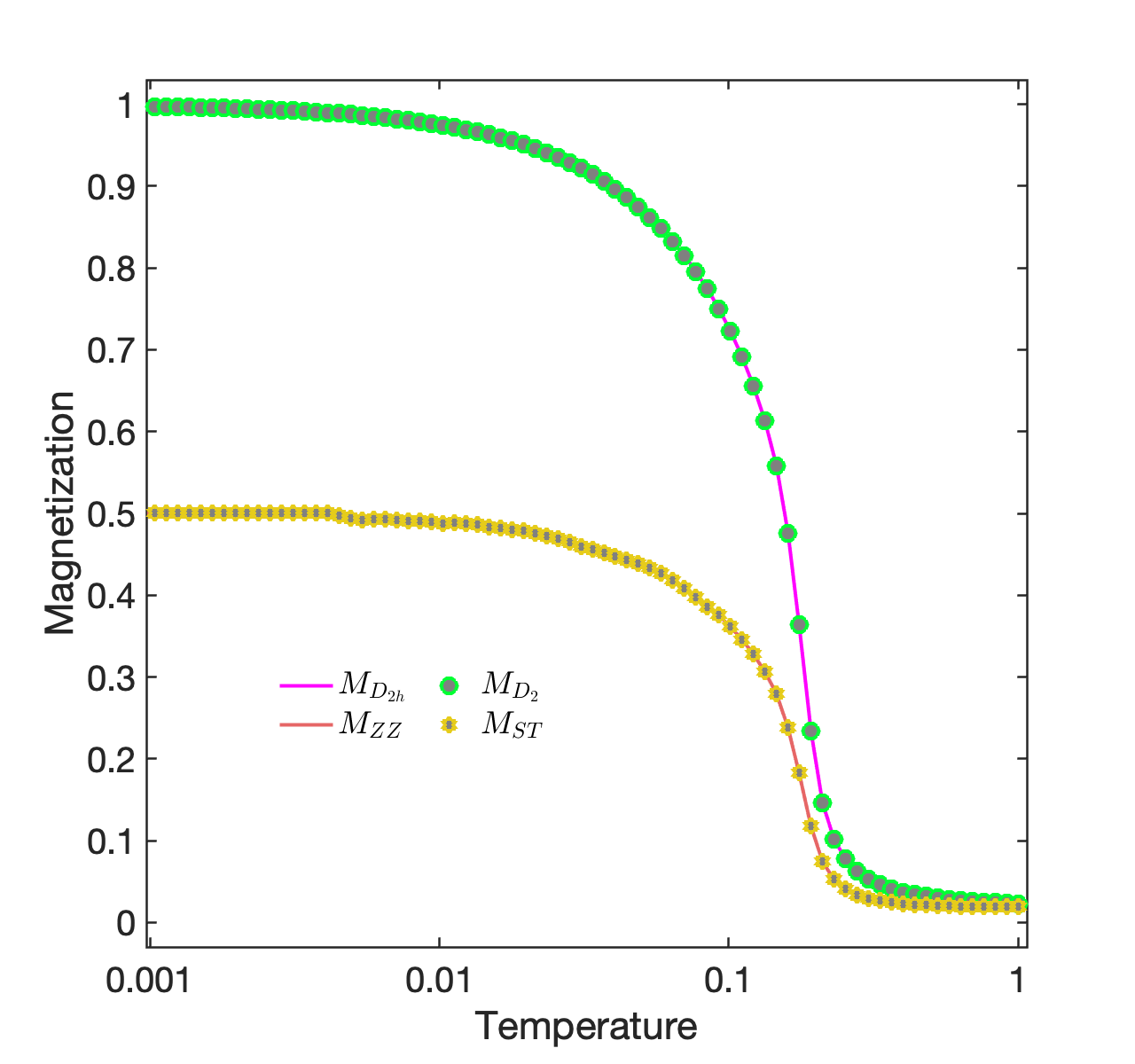}
  \caption{Magnetization as a function of temperature at the $O(3)$ points, with $\varphi \approx 0.65 \pi$ for the zigzag (ZZ) and $D_{2h}$ orders and $\varphi \approx 1.65 \pi$ for the stripy (ST) and $D_{2}$ orders. The $D_{2h}$ (ZZ) and $D_{2}$ (ST) curves show the same behavior as the Heisenberg-Kitaev model is symmetric under a sub-lattice transformation $J \rightarrow -J, K \rightarrow -K$, and meanwhile $S_i \rightarrow -S_i$ for either of the honeycomb sublattices.}
  \label{fig:mags_T}
\end{figure}

Consider a $Z$-type zigzag (stripy) moment for instance. As measured in Figure~\ref{fig:mags_T}, its expectation value at the $O(3)$ point is $M = \frac{1}{2}$ when $T\rightarrow 0$.
This can be understood by parametrizing the reference spin as 
$\vec{S}_{0} = (\sin\theta\sin\phi \ \sin\theta\cos\phi \ \cos\theta)^{\rm T}$, where $\theta, \phi$ are Euler angles.
Since spins in those states are actually arranged according to the $D_{2h}$ ($D_2$) pattern, the zigzag (stripy) moment of an individual sample is $\vec{m} = (0 \ 0 \ \cos\theta)^{\rm T}$, 
and the corresponding ensemble average, by integrating over all allowed states, is  $M = \frac{1}{4\pi}\int \abs{\vec{m}} \sin\theta d\theta d\phi = \frac{1}{2}$.
Hence, the $D_{2h}$ and $D_{2}$ orders provide a more universal and complete description for the magnetization as compared to the zigzag and stripy order.
(There is no phase transition or crossover separating the $O(3)$ points from the neighboring points at $T\rightarrow 0$.
For instance, it can be shown that the ground-state energy of the ZZ/$D_{2h}$ phase is $E = \frac{1}{3}(-K + J)$ per bond. This energy is degenerate with that of the Neel and FM order, $E_{\rm N} = - \frac{1}{3}(K + J)$ and $E_{\rm FM} = \frac{1}{3}(K + J)$, at $J = 0$ and $J = -K$, respectively, which are the two phase boundaries at $\varphi = \frac{\pi}{2}, \frac{3\pi}{4}$.
Within the ZZ/$D_{2h}$ regime, the ground-state energy is linear to $J/K$.
Nonetheless, TK-SVM is also capable of distinguishing states with continuous and discrete degeneracy; see Appendix~\ref{app:rank-2}.)

\section{Hidden $O(3)$ symmetry} \label{sec:hidden}
The $D_2$ and $D_{2h}$ ordering matrices in Table~\ref{tab:ops} comprise a finite set of orthogonal matrices, which preserve the spin length and are invertible.
This means that, by inverting those transformations, the $D_2$ and $D_{2h}$ order can be converted to simple ferromagnets.

Specifically, one can define spin orientations in a sublattice-dependent coordinate, $\tilde{S}_k = \hat{T}^{\scalebox{0.618}{A,B}}_k \vec{S}_k$.
The magnetization Eq.~\eqref{eq:op} then becomes 
$\overrightarrow{M} = \widetilde{M} = \sum_k \tilde{S}_k$, 
describing a ferromagnetic alignment of $\tilde{S}$ spins.

The above transformation acts on spin patterns. Naturally, one examines the form of the Hamiltonian in the same coordinate system.
Without loss of generality, we focus on the interaction of a local bond $\corr{kl}_\gamma$ , which can be rewritten as
\begin{align}\label{eq:H_local}
	H_{kl} = \vec{S}_k^{\rm \, T} \hat{J}_{\gamma} \vec{S}_l,
\end{align} 
where $\hat{J}_{\gamma}$ corresponds to the three types of bonds in the Hamiltonian Eq.~\eqref{eq:model} with $\gamma \in \{x, y, z\}$,
\begin{align} \label{eq:J}
	\begin{psmallmatrix}
		K+J & & \\
		 & J & \\
		 & & J
	\end{psmallmatrix}, \ 
		\begin{psmallmatrix}
		J & & \\
		 & K+J & \\
		 & & J
	\end{psmallmatrix}, \ 
	\begin{psmallmatrix}
		J & & \\
		 & J & \\
		 & & K+J
	\end{psmallmatrix}.
\end{align}

Under the sublattice-dependent coordinate transformations, Eq.~\eqref{eq:J} becomes 
\begin{align}\label{eq:H_local_sub}
	\widetilde{H}_{kl} = \tilde{S}_k^{\rm T} \hat{T}^{\scalebox{0.618}{A,B}}_k
				\hat{J}_{\gamma} \big(\hat{T}^{\scalebox{0.618}{A,B}}_{k}\big)^{\rm T} \tilde{S}_l.
\end{align} 
The three different bonds transform as
\begin{gather}
 \hat{J}_{x}	\rightarrow \big(\hat{T}^{\scalebox{0.618}{A,B}}_2\big)^{\rm T} \hat{J}_x \hat{T}^{\scalebox{0.618}{A,B}}_3, \ 
	\big(\hat{T}^{\scalebox{0.618}{A,B}}_4\big)^{\rm T} \hat{J}_x \hat{T}^{\scalebox{0.5}{A,B}}_1 \nonumber \\  \hat{J}_{y}	\rightarrow
	 \big(\hat{T}^{\scalebox{0.5}{A(B)}}_{3(1)}\big)^{\rm T} \hat{J}_y \hat{T}^{\scalebox{0.5}{B(A)}}_{1(3)}, \
	\big(\hat{T}^{\scalebox{0.5}{A(B)}}_{4(2)}\big)^{\rm T} \hat{J}_y \hat{T}^{\scalebox{0.5}{B(A)}}_{2(4)} \nonumber \\
	\hat{J}_{z}	\rightarrow 
	 \big(\hat{T}^{\scalebox{0.618}{A,B}}_1\big)^{\rm T} \hat{J}_z \hat{T}^{\scalebox{0.5}{A,B}}_2, \
	\big(\hat{T}^{\scalebox{0.5}{A,B}}_3\big)^{\rm T} \hat{J}_z \hat{T}^{\scalebox{0.5}{A,B}}_4, \nonumber
\end{gather}
respectively leading to
\begin{align}\label{eq:J_sub}
	\pm \begin{psmallmatrix}
		K+J & & \\
		 & -J & \\
		 & & -J
	\end{psmallmatrix}, \ 
	\pm \begin{psmallmatrix}
		-J & & \\
		 & K+J & \\
		 & & -J
	\end{psmallmatrix}, \ 
	\pm \begin{psmallmatrix}
		-J & & \\
		 & -J & \\
		 & & K+J
	\end{psmallmatrix},
\end{align}
where ``$+$'' (``$-$'') corresponds to the $D_2$ ($D_{2h}$) order.

Clearly, at $K = -2J$, the couplings in the sublattice coordinate reduce to isotropic matrices, $\pm J\mathbbm{1}$, where $\mathbbm{1}$ denotes the identity matrix.
$\widetilde{H}_{kl}$ is simply the local interaction for a ferromagnetic Heisenberg model of spin $\tilde{S}$, with $J > 0$ ($< 0$) in the $D_2$ ($D_{2h}$) phase.
This precisely reproduces the hidden $O(3)$ symmetries of the HK model, which were previously identified in Ref.~\cite{Chaloupka15} by a dual transformation.

The above way of identifying hidden symmetries is especially straightforward. It does not use specific properties and hence does not rely on prior insights of a Hamiltonian.
The high-symmetry points are self-evident once the order parameters are detected.
Importantly, as shown in Figure~\ref{fig:pd} (b), data from the high-symmetry points are \emph{not} needed in the training.

\begin{figure}[t]
  \centering
  \includegraphics[width=0.45\textwidth]{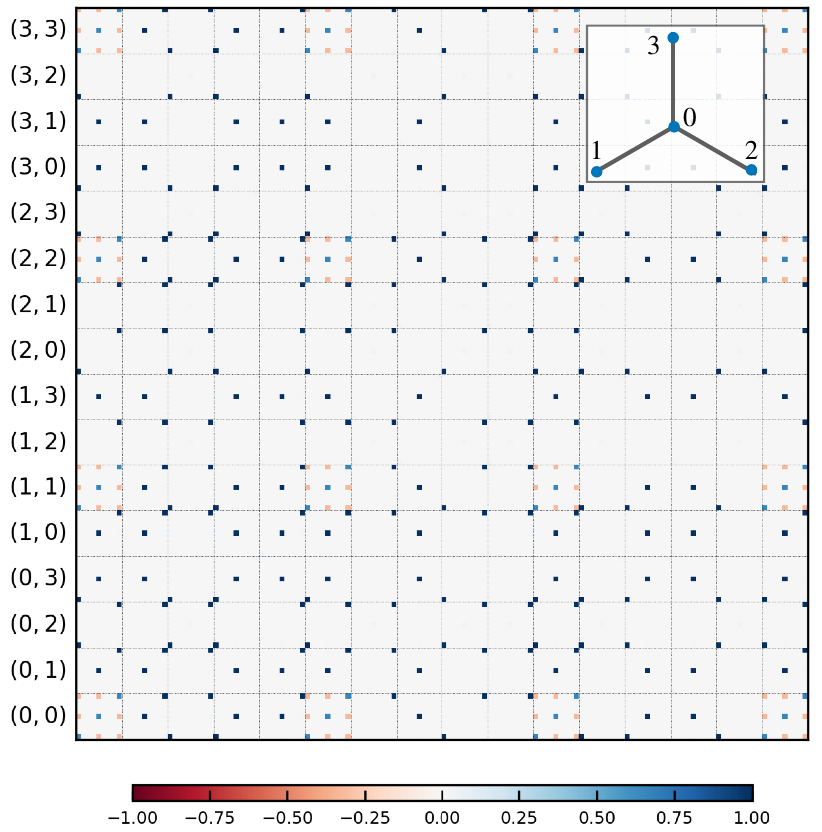}
  \caption{The $C_{\mu\nu}$ matrix learned by a rank-$2$ TK-SVM with a four-spin triad cluster (inner panel) at the boundary point $\varphi = \frac{3\pi}{4}$. The axes iterate over spin indices $(i, j)$ and spin components $(\alpha, \beta)$ in a lexicographically order, from bottom (left) to top (right).
  The spin indices divide the $C_{\mu\nu}$ matrix into $9\times 9$ subblocks. Non-vanishing entries in a block represent the form of correlations between quadratic components $S_i^\alpha S_j^{\beta}$ and $S_{i^\prime}^{\alpha^\prime} S_{j^\prime}^{\beta^\prime}$. Blocks with $i=j$ and $i^\prime = j^\prime$ lead to constants owing to the trivial normalization $\abs{\vec{S}} = 1$. Other blocks corresponds to the local constraints $G_1$ and $G_2$.
  The pattern learned for $\varphi = \frac{7\pi}{4}$ (not shown) has a similar structure with sign flips in certain entries.} 
  \label{fig:C_phi9}
\end{figure}

 \begin{figure}[t]
 	\begin{tabular}{c}
 		\subfloat[$\varphi=\frac{3\pi}{4}$]{\includegraphics[width=0.34\textwidth]{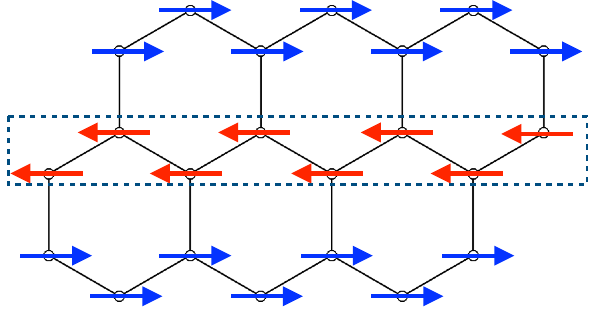}} \\
 		\subfloat[$\varphi=\frac{7\pi}{4}$]{\includegraphics[width=0.34\textwidth]{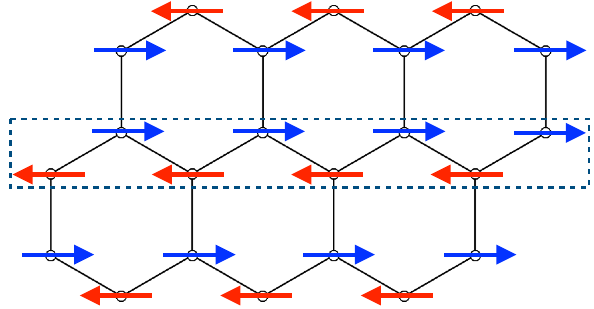}} 
 	\end{tabular}
 \caption{Representative classical ground-state configuration at $\varphi = \frac{3\pi}{4}, \frac{7\pi}{4}$. The system forms ferromagnetic (a) or anti-ferromagnetic (b) Ising chains. The subdimensional symmetry leads to a classical subextensive degeneracy by flipping one entire chain of spins.}
 \label{fig:subsymmetry}
 \end{figure}

\section{Local constraints at phase boundaries}\label{sec:constrains}
In general, at a phase boundary, the competition between the orders of the two phases can lead to more subtle properties such as an enhanced symmetry or an (emergent) local constraint. 
In this section we discuss the local constraints learned at the phase boundaries in the phase diagram of Figure~\ref{fig:pd}.  The cases $\varphi = \frac{\pi}{2}$ and  $\frac{3\pi}{2}$ correspond to pure Kitaev models and are not discussed here further because  we already know from Ref.~\cite{Liu21}  that TK-SVM is able to learn the ground-state constraints for classical Kitaev spin liquids.
We focus here therefore on the boundaries at $\varphi = \frac{3\pi}{4}$ and $ \frac{7\pi}{4}$.

In general, a rank-$n$ TK-SVM detects rank-$n$ tensorial orders and correlations~\cite{Greitemann19, Liu19}.
To detect the local constraints, a rank-$2$ TK-SVM detecting quadratic correlations needs to be used.
Figure~\ref{fig:C_phi9} shows the rank-$2$ $C_{\mu\nu}$ matrix for $\varphi = \frac{3\pi}{4}$. We refer to our previous works Refs.~\cite{Liu19, Greitemann19b} for the systematic decoding of such matrix.
 The pattern for $\varphi = \frac{7\pi}{4}$ has a similar structure but displays different signs for certain entries.

\begin{figure}[t]
  \centering
  \includegraphics[width=0.45\textwidth]{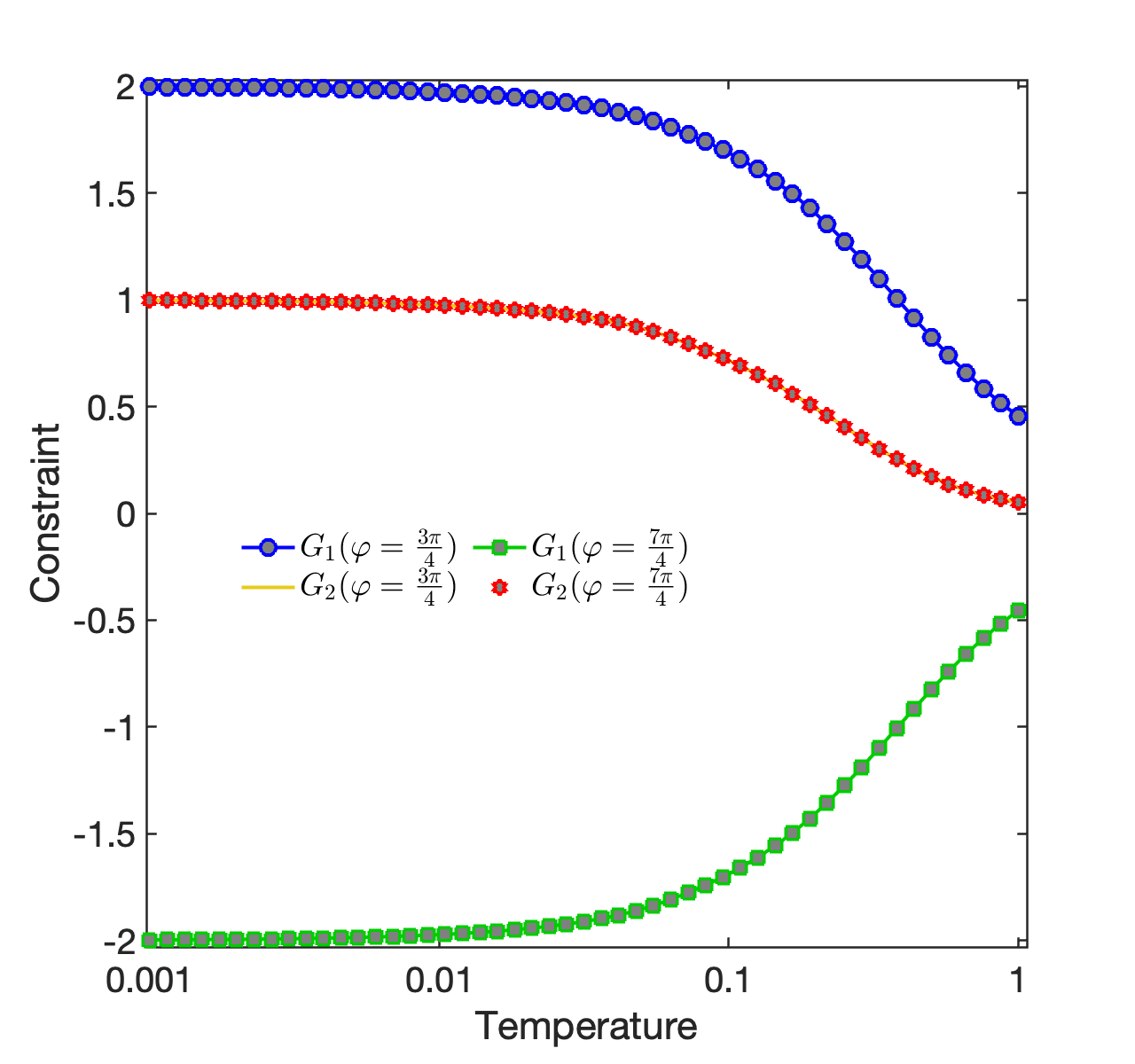}
  \caption{Local constraints at $\varphi = \frac{3\pi}{4}, \frac{7\pi}{4}$ as a function of temperature. $G_1$ and $G_2$ satisfy Eqs.~\eqref{eq:G1} and~\eqref{eq:G2} in the  ground state. (The $G_2$ curves at the two $\varphi$ values overlap.) }
  \label{fig:Gs_T}
\end{figure}

Two constraints, $G_1$ and $G_2$, are inferred,
\begin{align}
	& G_{1}\!  = \! \langle S^{x}_0 (S^{x}_2 \! + \! S^{x}_3)  + S^{y}_0 (S^{y}_1 \! + \! S^{y}_3) + S^{z}_0 (S^{z}_1 \! + \! S^{z}_2)\rangle_{\rm td} = \pm 2, \label{eq:G1}\\
	& G_{2} = \langle S^{x}_2 S^{x}_3 + S^{x}_1 S^{x}_3 + S^{x}_1 S^{x}_2 \rangle_{\rm td} = 1, \label{eq:G2}
\end{align}
with all other nearest neighbor and next-nearest neighbor correlations vanishing.
Here, $\langle . \rangle_{\rm td}$ denotes a lattice average over triad clusters involving three bonds and four spins (see the inner panel of Figure~\ref{fig:C_phi9}), and ``$+$", ``$-$" correspond to $\varphi = \frac{3\pi}{4}, \frac{7\pi}{4}$, respectively.
These constraints are verified by their explicit measurement in a Monte Carlo simulation as shown in Figure~\ref{fig:Gs_T}.

The local constraints $G_1$ and $G_2$ are invariant under the following transformations,
\begin{gather}
	S^x_0,\, S^x_2, \, S^x_3 \rightarrow -S^x_0, \, -S^x_2, \, -S^x_3, \\
	S^y_0, \, S^y_3,\,  S^y_1 \rightarrow -S^y_0, \, -S^y_3, \, -S^y_1, \\
	S^z_0,\, S^z_1, \, S^z_2 \rightarrow -S^z_0, \, -S^z_1, \, -S^z_2. \label{eq:subsymmetry_z}
\end{gather}
However, since a spin is shared by two triads, these do not define a local, but rather a \emph{subdimensional} symmetry.
For instance, Eq.~\eqref{eq:subsymmetry_z} corresponds to a transformation flipping the $S_z$ component of spins in a chain formed by $x$- and $y$-bonds, as depicted in Figure~\ref{fig:subsymmetry}.

The solutions of Eqs.\eqref{eq:G1} and~\eqref{eq:G2} give the classical ground states.
The absence of cross terms, such as $S_i^{\alpha}S_j^{\beta \neq \alpha}$, indicates that each spin has only a single non-vanishing component in the ground state.
To satisfy the two constraints, the system thereby forms ferromagnetic ($\varphi = \frac{3}{4}\pi$) and anti-ferromagnetic ($\varphi = \frac{7}{4}\pi$) Ising chains.
Owing to the subdimensional symmetry, it does not cost energy to flip one Ising chain, leading to a subextensive line degeneracy with $3 \times 2^{L}$ classical ground states.
In other words, a ZZ/$D_{2h}$ (ST/$D_2$) order is degenerate with a FM (Neel) order at these boundary points.

In the spin-$\frac{1}{2}$ HK model, this subextensive degeneracy will be lifted by quantum fluctuations via a quantum order-by-disorder mechanism~\cite{Chaloupka10,Chaloupka13}.
Nevertheless, from the point of view of machine learning, the above application implies a possibility of using machine-learning approaches to generate non-trivial spin models.
The constraint $G_1$ is essential for the HK Hamiltonian at the classical phase boundaries where $J=-K$.
However, during training, the Hamiltonian is \emph{not} given to the machine but rather learned from the spin configurations.
Hence one can consider potential applications to learn non-trivial spin Hamiltonians from samples of simple orders.

\section{Summary and outlook}\label{sec:conclusion}
In summary, we demonstrated that TK-SVM provides a data-driven approach to the problem of identifying hidden symmetries in phases with unconventional magnetic orders.
In comparison with other constructions, which are typically  contingent on the skill and experience of the researcher, this approach does not require particular knowledge of the Hamiltonian and is feasible even when prior insight in the system is limited.

We considered the honeycomb Heisenberg-Kitaev model as an example and successfully identified its hidden $O(3)$ points and the associated transformations.
We also clarified that the $D_{2h}$ and $D_{2}$ orders provide a more universal description of the magnetization compared to  zigzag and stripy order.
Our results emphasize the significance of being able to express the order parameter explicitly in  many-body spin systems, which can be done by an interpretable machine-learning method like TK-SVM.

Moreover, we showed that our machine is also capable of revealing subdimensional symmetries.
On the one hand, this complements our previous study of Ref.~\cite{Liu21} which showed that TK-SVM identified the local $Z_2$ symmetry of classical Kitaev spin liquids by probing their ground-state constraints.
On the other hand, as such symmetries are typically related to degenerate competing orders, their identification by machine learning methods implies a potential generative use of these machines.
One could consider applications to learn non-trivial spin Hamiltonians from moderate datasets of simple orders and use the learned Hamiltonians to further generate more interesting phases. 

Hidden symmetries are also found in symmetry-protected topological states~\cite{Gu09, Chen11, Pollmann12} such as the hidden $Z_2 \times Z_2$ symmetry in the celebrated Haldane phase~\cite{Affleck87, Affleck88, Kennedy92, Kennedy92b}.
The Haldane phase, as well as an array of other symmetry-protected topological states, can be mapped onto Landau-type orders by a nonlocal unitary transformation associated with the respective hidden symmetry~\cite{Kennedy92, Kennedy92b, denNijs89, Tu08, Tu08b, Tu09, Else13, Duivenvoorden13}.
How to detect such hidden symmetries with machine-learning techniques is an interesting topic left for future work.
While it might be easier to construct an {\it ad hoc} machine for such a particular SPT phase, devising a versatile machine that is applicable to a (reasonably) wide class of topological phases remains however a challenging task.

\section*{Open source and Data availability}
The TK-SVM library has been made openly available with documentation and examples~\cite{Jonas}.
The data used in this work are available upon request.

\begin{acknowledgements}
We thank Philippe Corboz, Matthias Gohlke, Jheng-Wei Li, Hao Song and Hong-Hao Tu  for useful discussions.
NR, KL, and LP acknowledge support from FP7/ERC Consolidator Grant QSIMCORR, No. 771891, and the Deutsche Forschungsgemeinschaft (DFG, German Research Foundation) under Germany's Excellence Strategy -- EXC-2111 -- 390814868.
Our simulations make use of the $\nu$-SVM formulation~\cite{Scholkopf00}, the LIBSVM library~\cite{Chang01, Chang11}, and the ALPS\-Core library~\cite{Gaenko17}.
\end{acknowledgements}

\begin{appendix}

\section{Setting up of TK-SVM } \label{app:svm}
\begin{figure}[t]
\centering
 	\begin{tabular}{c}
 		\subfloat[$D_{2}$]{\includegraphics[width=0.4\textwidth]{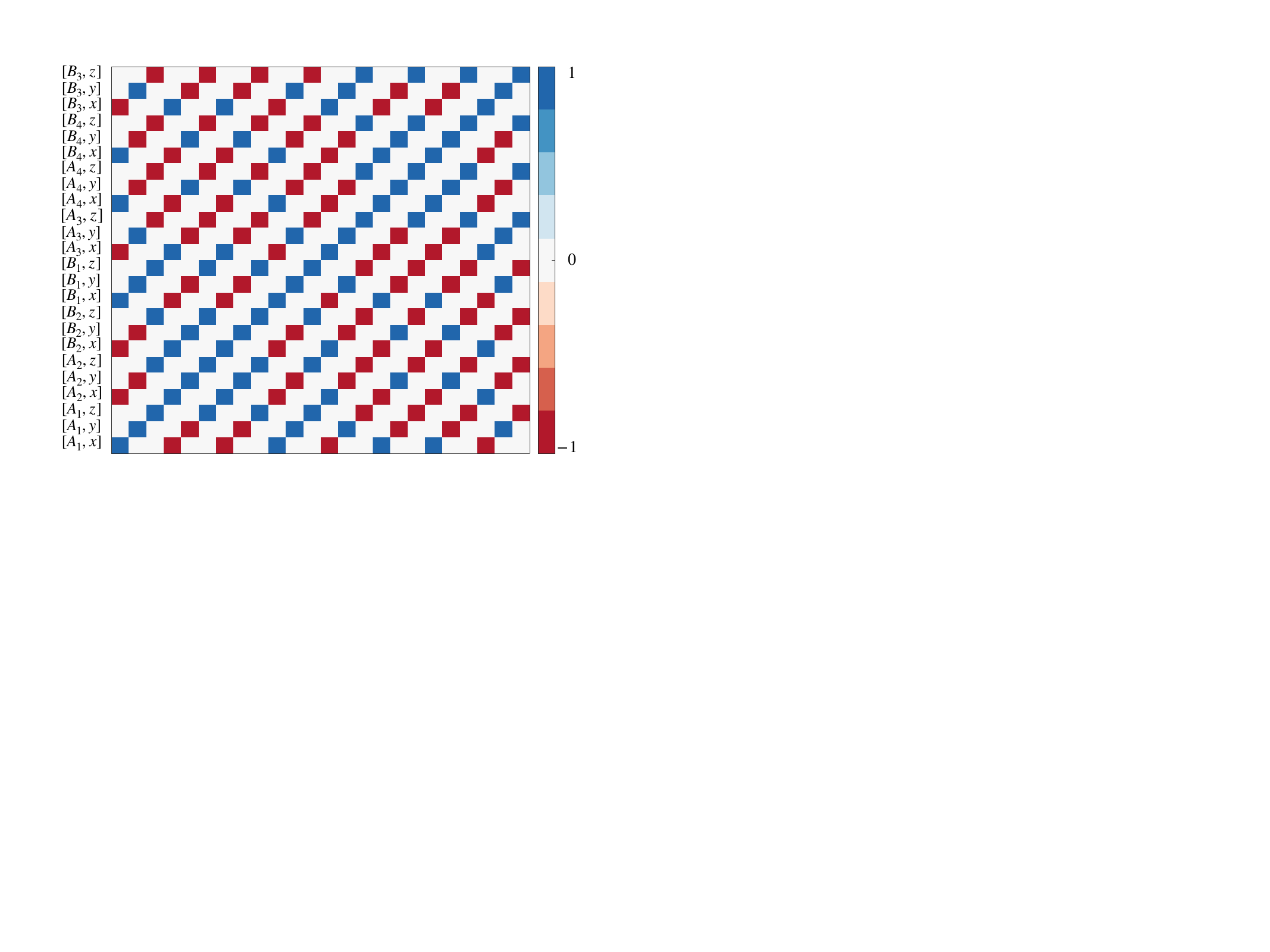}} \\
 		 \subfloat[$D_{2h}$]{\includegraphics[width=0.4\textwidth]{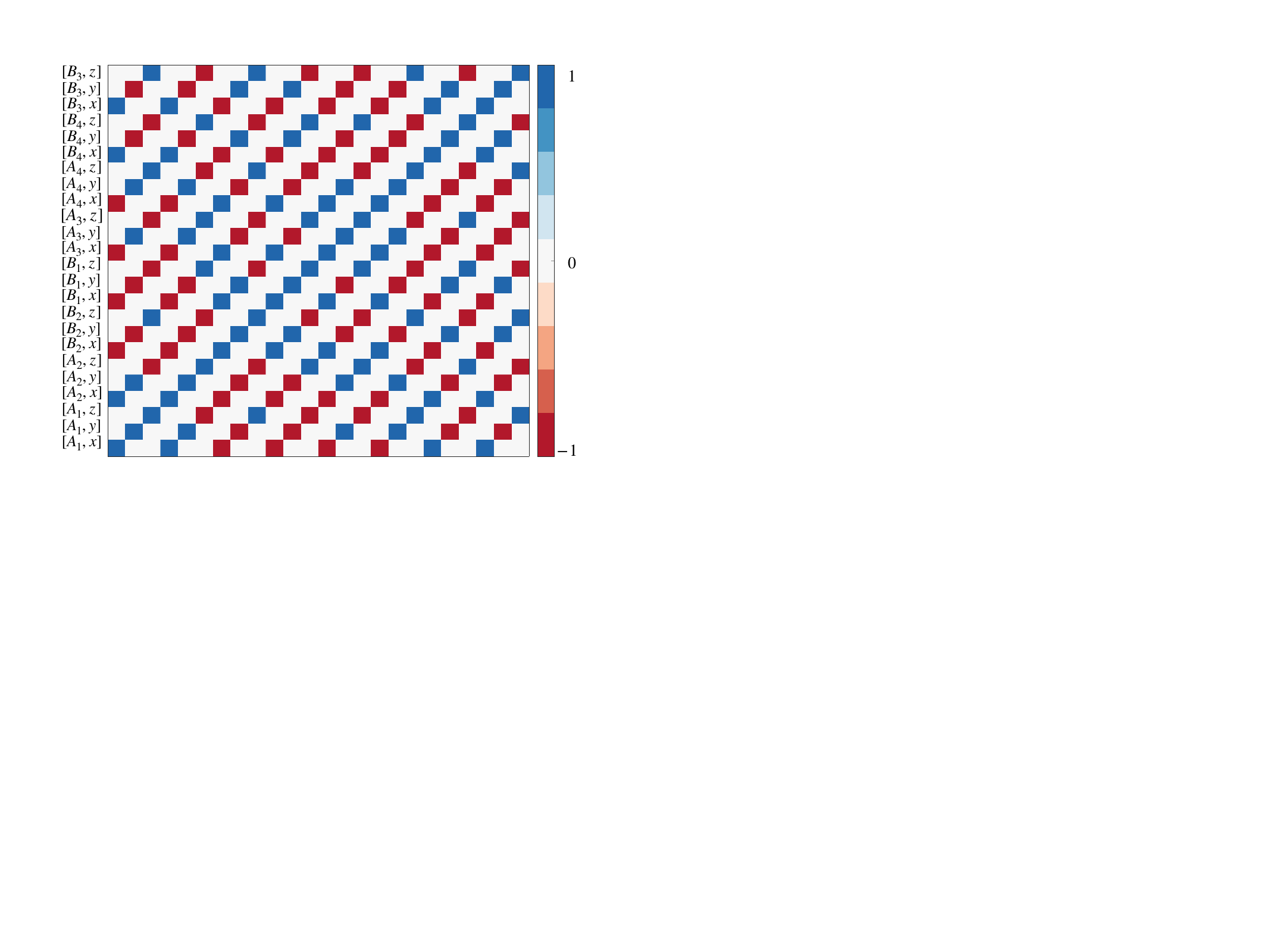}} 
 	\end{tabular}
 \caption{The $C_{\mu\nu}$ matrices learned by a rank-$1$ TK-SVM in the ST/$D_2$ and ZZ/$D_{2h}$ phases. Each entry represents a correlation between two spin components defined by the weighted sum of the support vectors. Results of an eight-spin cluster ($2\times 2$ honeycomb unit cells), which is the minimal unit of the $D_2$ and $D_{2h}$ order, are shown for demonstration. From bottom to top, the vertical axis is labeled in the same convention as in the lattice Figure~\ref{fig:lattice}. The same labeling applies to the horizontal axis from left to right. The interpretation of these patterns leads to the respective ordering matrices in Table~\ref{tab:ops}.}
 \label{fig:C_munu}
 \end{figure}

Here we provide more details of TK-SVM and refer the reader to Refs.~\cite{Greitemann19, Liu19} for the introduction of the method and Ref.~\cite{Greitemann19b} for a review, including comprehensive discussions on how to interpret $C_{\mu\nu}$ matrices.

The map $\phi$ in the decision function Eq.~\eqref{eq:d(x)} maps a spin sample $\mathbf{x}$ to a configuration of degree $n$ monomials,
\begin{align}\label{eq:phi}
    \phi : \mathbf{x} \rightarrow \phi(\mathbf{x} ) =
     \{ \phi_{\mu} \} = \{\langle S^{\alpha_{1}}_{a_{1}}... S^{\alpha_{n}}_{a_{n}}\rangle_{\rm cl} \},
\end{align}
where $n$ also corresponds to the rank of a TK-SVM.
This mapping partitions the system into clusters containing \textit{r} spins labeled with $\alpha_{n} = \{1 , 2, \dots, r\}$, while
$\mu = \{\alpha_{n} , a_{n}\} = \{\alpha_{1}, a_{1}, ..., \alpha_{n}, a_{n}\}$ denotes a collective index.
Then a cluster average $\langle . \rangle_{\rm cl}$ is introduced for dimension reduction.
This construction of feature vectors makes use of the fact that local orders and local constraints can be generally expressed by a finite number of spins.
In potential extensions to quantum systems, such construction may still be done to detect local order parameters.
The cluster average is not suitable for non-local orders.
Nevertheless, in cases in which a system can be characterized by short-ranged entanglement, one may consider using local correlators sampled from a $k$-particle reduced density matrix to construct the feature space.

The optimal choice for the size and shape of the clusters in Eq.~\eqref{eq:phi} is in general unknown \emph{a priori}, and different phases in a phase diagram may have distinct translational symmetries.
Therefore, in practice, we adopt clusters comprising a large number of lattice unit cells in order to accommodate diverse orders.
In the results presented in the current paper, clusters with a size up to $288$ spins ($12 \times 12$ honeycomb unit cells) were used.

The $C_{\mu\nu}$ matrix is defined by a weighted sum over support vectors,
\begin{align}\label{eq:C_munu}
    C_{\mu \nu} = \sum_{k} \lambda_{k} \phi_{\mu}\big(\mathbf{x}^{(k)}\big) \phi_{\nu} \big(\mathbf{x}^{(k)}\big),
\end{align}
where $\lambda_{k}$ is a Lagrange multiplier with $\lambda_k \neq 0$ corresponding to support vectors, and non-vanishing entries of $C_{\mu\nu}$ represent correlations between particular monomial components.
Standard SVM optimizations, which maximizes the separating margin~\cite{BookVapnik}, are employed to solve $\lambda_{k}$.  
We refer to Ref.~\onlinecite{Liu19} for concrete formulations of the optimization problem and the construction of the kernel.

The $C_{\mu\nu}$ matrices learned in the ST/$D_2$ and ZZ/$D_{2h}$ phases are shown in Figure~\ref{fig:C_munu} for example.
The alternating colors indicate sign flips on individual spin components.
The corresponding order parameters are given in Table~\ref{tab:ops}, and the systematic procedure of decoding $C_{\mu\nu}$ matrices can be found in Refs.~\cite{Liu19, Greitemann19b}.

\begin{figure*}[t]
    \centering
    \includegraphics[width=1\textwidth]{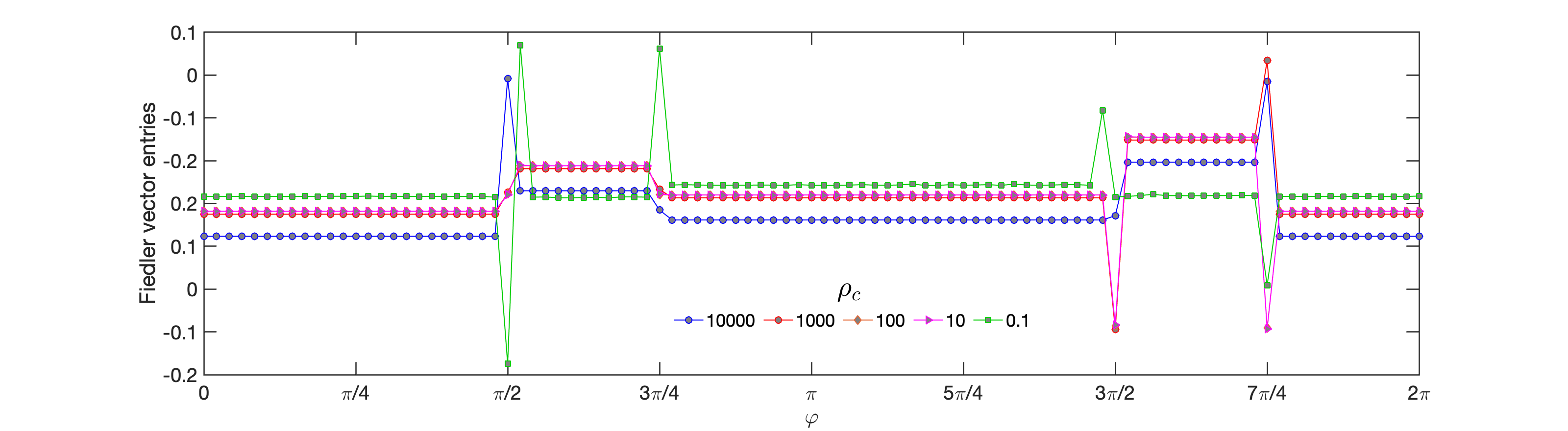}
    \caption{Fiedler vectors obtained with different choices of $\rho_c$. 
    	In all cases, where $\rho_c$ is large enough to set a characteristic scale ``$\gg 1$'' for the reduced $\rho$ criterion Eq.~\eqref{eq:rho_rules}, the clustering is evident and robust.
    	The profound jumps at $\varphi = \frac{\pi}{2}, \, \frac{3\pi}{4}, \, \frac{3\pi}{2}, \, \frac{7\pi}{4}$ correspond to phase boundaries, as they do not belong to any plateaus (stable phases). 
    	A case of small $\rho_c = 0.1$ is also included for comparison.
    	$\rho_c = 100$ is used in the maintext.}
    \label{fig:comparing_FV}
\end{figure*}

\begin{figure}
    \centering
    \begin{tabular}{cc}
    	\subfloat[]{\includegraphics[width=0.24\textwidth]{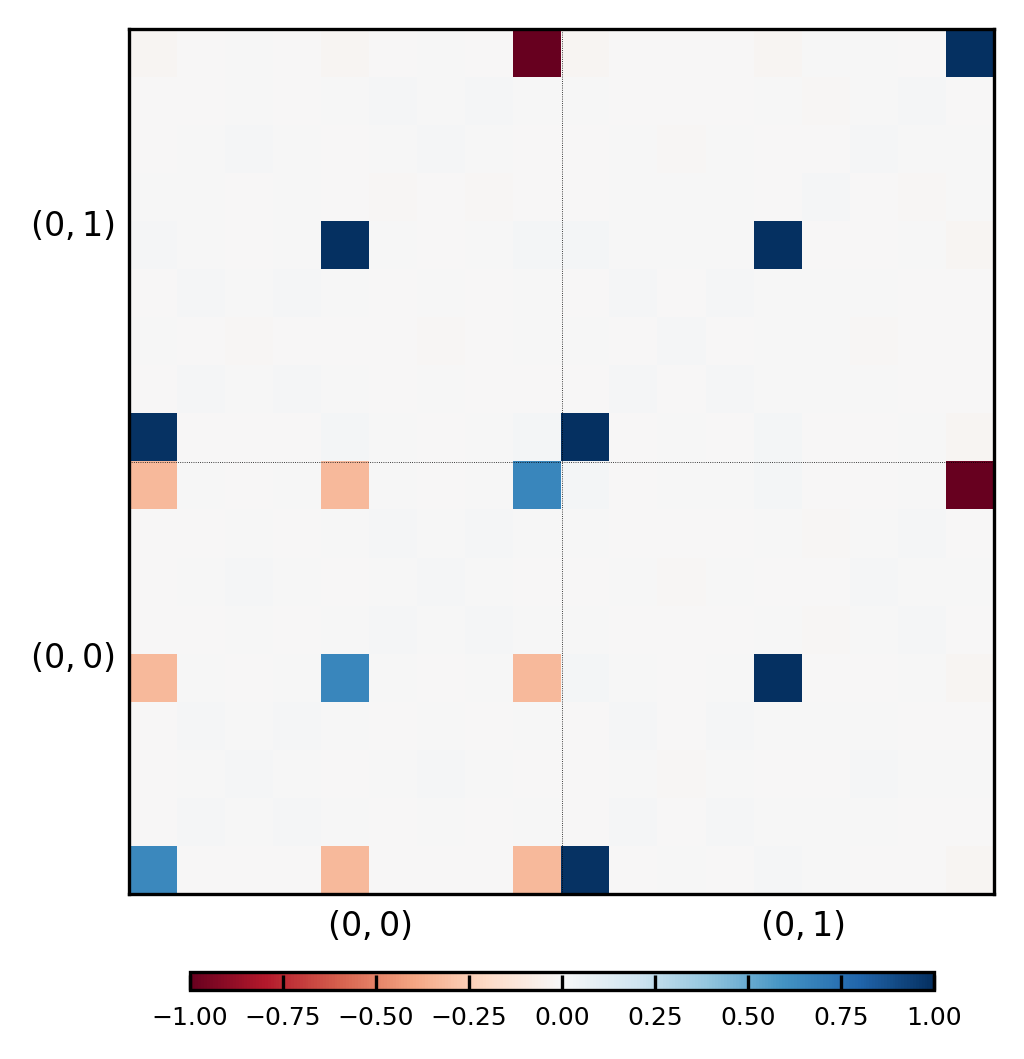}} &
    	\subfloat[]{\includegraphics[width=0.24\textwidth]{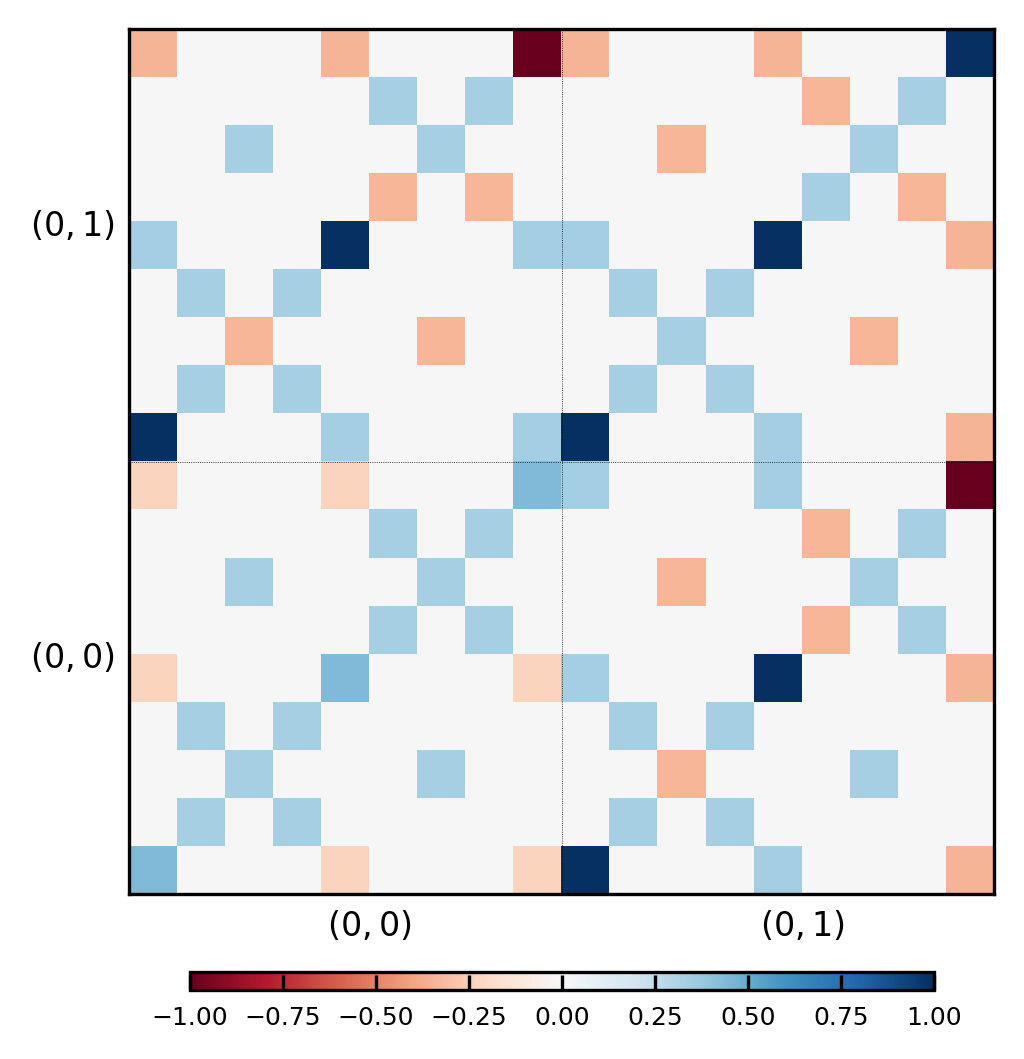}}
    \end{tabular}
    \caption{Representative blocks of the $C_{\mu\nu}$ matrices of the ZZ/$D_{2h}$ phase learned by a rank-$2$ TK-SVM with the eight-spin $D_{2h}$ magnetic cell, away from (a) and at (b) the $O(3)$ point. Blocks are labeled by the spin indices $(i,j)$. Nonvanishing entries in a block correspond to correlations between quadratic components $S_i^\alpha S_j^{\beta}$ and $S_{i^\prime}^{\alpha^\prime} S_{j^\prime}^{\beta^\prime}$. Negative elements in the $(0,0)$ block reflect the spin normalization $\abs{\vec{S}} = 1$. Non-trivial entries in (a) are the diagonal ones in each $9 \times 9$ subblock.}
    \label{fig:ZZ_rank2}
\end{figure}

\section{Details of Graph Partitioning} \label{app:graph}
For a binary classification between two sample sets ``A'' and ``B'', the parameter $\rho$ in decision function Eq.~\eqref{eq:d(x)} behaves as
\begin{align}\label{eq:rho_rules}
    \abs{\rho_{\rm AB}} \begin{cases}
        \gg 1 & \textup{${\rm A, B}$ in the same phase}, \\
        \lesssim 1 & \textup{${\rm A, B}$ in different phases},
    \end{cases}
\end{align}
which is referred to as the reduced $\rho$ criterion~\cite{Liu19, Greitemann19b}.

The weight of an edge is determined by $\rho$ in the decision function learned for the two end points, with a Lorentzian weighting function, 
\begin{align}\label{eq:weight}
  w(\rho) &= 1 - \frac{\rho_c^2}{(|\rho|-1)^2+\rho_c^2} \in [0,1),
\end{align}
where $\rho_c$ is a super-parameter introduced to set a characteristic scale for ``$\gg 1$" in the above reduced $\rho$ criterion.
However, as we will show in Figure~\ref{fig:comparing_FV}, the choice of $\rho_c$ is not crucial.

The graph can be described by a Laplacian matrix,
\begin{align}\label{eq:L_mat}
  \hat{L} = \hat{D} - \hat{A} =
   \begin{bmatrix}
d_{1} & -w_{12} & ... & -w_{1M} \\ 
 -w_{21}& d_{2} & ... & -w_{2M}\\ 
 \vdots &  &  & \vdots \\ 
- w_{M1}& -w_{M2} & ... & d_{M}
\end{bmatrix}.
\end{align}
Here, the off-diagonal entries, 
$\omega_{ij} = \omega_{ji} = \omega(\rho_{ij})$,
host all the edge weights and are collected by the adjacency matrix $\hat{A}$.
The diagonal entries, $d_i = \sum_{j \neq i} \omega(\rho_{ij})$, represent degrees of the vertices and form the degree matrix $\hat{D}$.
$\hat{L}$ is symmetric by construction as only the magnitude of $\rho$ is used.
(The sign of $\rho$ can reveal which data set is more disordered, but this property is not needed for the graph partitioning; see Refs.~\cite{Liu19} and ~\cite{Greitemann19b} for details.) 
According to Fiedler's theory~\cite{Fiedler73, Fiedler75}, partitioning of a graph can be formulated as an eigenproblem of $\hat{L}$, as shown in Eq.~\eqref{eq:L_f}.
The second smallest eigenvector, known as the Fiedler vector, reflects the clustering of the graph.

In Figure~\ref{fig:comparing_FV}, we compare the resultant Fiedler vectors using different values of $\rho_c$.
The $M$ vertices are classified into distinct subgraph components (indicated by the plateaus).
In the case of $\rho_c = 0.1$, which does not suffice to define a scale ``$\gg 1$'', the partitioning is less obvious as all Fiedler vector entries display very similar values.
However, in all other cases, where $\rho_c$ crosses several orders, the clustering is clear and robust.

\section{Discriminating states in the same phase with different degeneracies}\label{app:rank-2}
In Section~\ref{sec:op} we discussed that there is no singularity separating the high-symmetry points with a continuous $O(3)$ degeneracy from their neighboring points with a discrete three-fold degeneracy.
Nevertheless, while they are thermodynamically in the same phase, those points can still be distinguished with the framework of TK-SVM.
This may be done by a rank-$2$ TK-SVM, as a magnetic order will also give finite quadratic correlations (which can be viewed as a ``redundant'' representation of the order parameter for the case at hand.)
Results for the ZZ/$D_{2h}$ phase are depicted in Figure~\ref{fig:ZZ_rank2} for instance.
The rank-$2$ $C_{\mu\nu}$ pattern away from the high-symmetry point displays nontrivial quadratic correlations between only the diagonal elements in each sub-block. The absent correlations between cross terms like $S^{\alpha}_iS^{\beta\neq \alpha}_j$ reflect locking of the spin orientation with a lattice axis, whereas such correlations are present in the pattern learned at the $O(3)$ point.
\end{appendix}

\bibliographystyle{apsrev4-1}
\bibliography{jk}

\end{document}